\documentclass[aps,prm,reprint,amsmath,amssymb,superscriptaddress]{revtex4-2}
\usepackage{graphicx}
\usepackage{dcolumn}
\usepackage{bm}
\usepackage{subfig} 
\usepackage{xcolor} 

\begin{document}

\title{Terahertz spin-current transparency through rough interfaces}

\author{Jiří Jechumtál}
\affiliation{Faculty of Mathematics and Physics, Charles University, 121 16 Prague, Czech Republic}

\author{Jakub Zázvorka}
\affiliation{Faculty of Mathematics and Physics, Charles University, 121 16 Prague, Czech Republic}

\author{Ondřej Novák}
\affiliation{Faculty of Mathematics and Physics, Charles University, 121 16 Prague, Czech Republic}

\author{Martin Rejhon}
\affiliation{Faculty of Mathematics and Physics, Charles University, 121 16 Prague, Czech Republic}

\author{Peter Kubaščík}
\affiliation{Faculty of Mathematics and Physics, Charles University, 121 16 Prague, Czech Republic}

\author{Lukáš Nowak}
\affiliation{Faculty of Mathematics and Physics, Charles University, 121 16 Prague, Czech Republic}

\author{Petr Němec}
\affiliation{Faculty of Mathematics and Physics, Charles University, 121 16 Prague, Czech Republic}

\author{Eva Schmoranzerová}
\affiliation{Faculty of Mathematics and Physics, Charles University, 121 16 Prague, Czech Republic}

\author{Martin Veis}
\affiliation{Faculty of Mathematics and Physics, Charles University, 121 16 Prague, Czech Republic}

\author{Lukáš Nádvorník}
\affiliation{Faculty of Mathematics and Physics, Charles University, 121 16 Prague, Czech Republic}

\author{Zdeněk Kašpar}
\affiliation{Faculty of Mathematics and Physics, Charles University, 121 16 Prague, Czech Republic}
\affiliation{Institute of Physics, Czech Academy of Science, 162 00 Prague, Czech Republic}

\date{\today}

\begin{abstract}

Spin transport across interfaces is critical for spintronic devices, yet remains difficult to probe on ultrafast timescales. We use terahertz emission spectroscopy on Co$|$Pt heterostructures whose interface roughness is tuned through the thickness of an underlying Au buffer layer, while leaving other growth parameters unchanged. From the measured THz electric field, we extract the interface spin-current transparency $t_s$ after correcting for the changes in sample impedance and optical absorption of the stack. Surprisingly, we find that $t_s$ decreases by only $\sim$30\% as the interface root-mean-square roughness and the lateral grain size both increase by a factor of three, with no measurable change in the THz spectrum. These results demonstrate that interfacial spin transport is relatively robust against morphological variations on ultrafast timescales, establishing terahertz emission spectroscopy as a reliable probe of spin dynamics across imperfect interfaces.

\end{abstract}

\maketitle

\section{Introduction}

Interfacial properties are the key parameters in proof-of-concept \cite{Wunderlich2010, Wang2022, Kang2025} as well as industrially produced \cite{Nguyen2024} spintronic devices, where they directly affect spin-pumping efficiency. A central role is played by the interface spin transparency $t_s$, defined as the fraction of spin current that successfully traverses the interface. This quantity is reduced by mechanisms such as interfacial spin memory loss \cite{Amin2016}. Understanding and controlling $t_s$ is therefore essential for device performance and for the development of new spintronic concepts. 

While $t_s$ has been extensively studied in the DC and GHz regimes \cite{Amin2016, Gu26, Jin2019, Rojas2014, Gupta2020kalu, Zhu19B, mazin2011} and partially by all-optical methods  \cite{Mukhopadhyay2023, Panda2019}, much less is known about how interface properties influence spin transport at THz frequencies. This becomes particularly relevant given the ongoing shift of spintronics toward ultrafast time scales, especially by the rapid advancements in antiferromagnetic \cite{Jungwirth2016, Baltz2018} and altermagnetic  \cite{Smejkal2022} spintronics. To experimentally probe the ultrafast spin currents and their penetration through the interface, the spintronic terahertz emitter (STE) \cite{Seifert2016} provides an ideal tool \cite{Jechumtal2024}. During the rapid development of STEs, and the simultaneous deeper understanding of the underlying physics \cite{Seifert2022}, many studies focused on improving  STE performance by modifying the interface properties \cite{Nenno2019, Li2019, Wang2023, Scheuer2022, Torosyan2018}. These studies, however, simultaneously modified multiple parameters (such as growth conditions, interlayer composition, defect density), making it difficult to isolate the role of interface morphology itself. As a result, an experiment isolating solely the role of interface roughness on spin transparency in the ultrafast regime is still missing.

In this work, we investigate the terahertz (THz) pulse emission originating in sub-picosecond spin-current pumping through a ferromagnet$|$heavy metal (FM$|$HM) interface of variable roughness. By analyzing the THz pulse strength, sample conductance and optical absorptance of the FM layer, we can directly see the role of interface roughness on changing spin transparency $t_s$. We observe no dependence of the THz pulse spectrum on interface roughness and only a $\sim$30\% decrease of $t_s$ despite the roughness root mean square (RMS) and the equivalent disk radius (EDR) varying by roughly a factor of 3. 

This finding is favorable for THz spintronic devices, since spin transparency remains relatively robust against a significant increase in interface roughness. Highly efficient operation can therefore be maintained without intensive interface engineering, while still leaving room for fine-tuning through morphological control.

\section{Experimental Methods}

\subsection{Spintronic THz emission and spin transparency extraction}

\begin{figure}[t]
\centering
\includegraphics[width=\columnwidth]{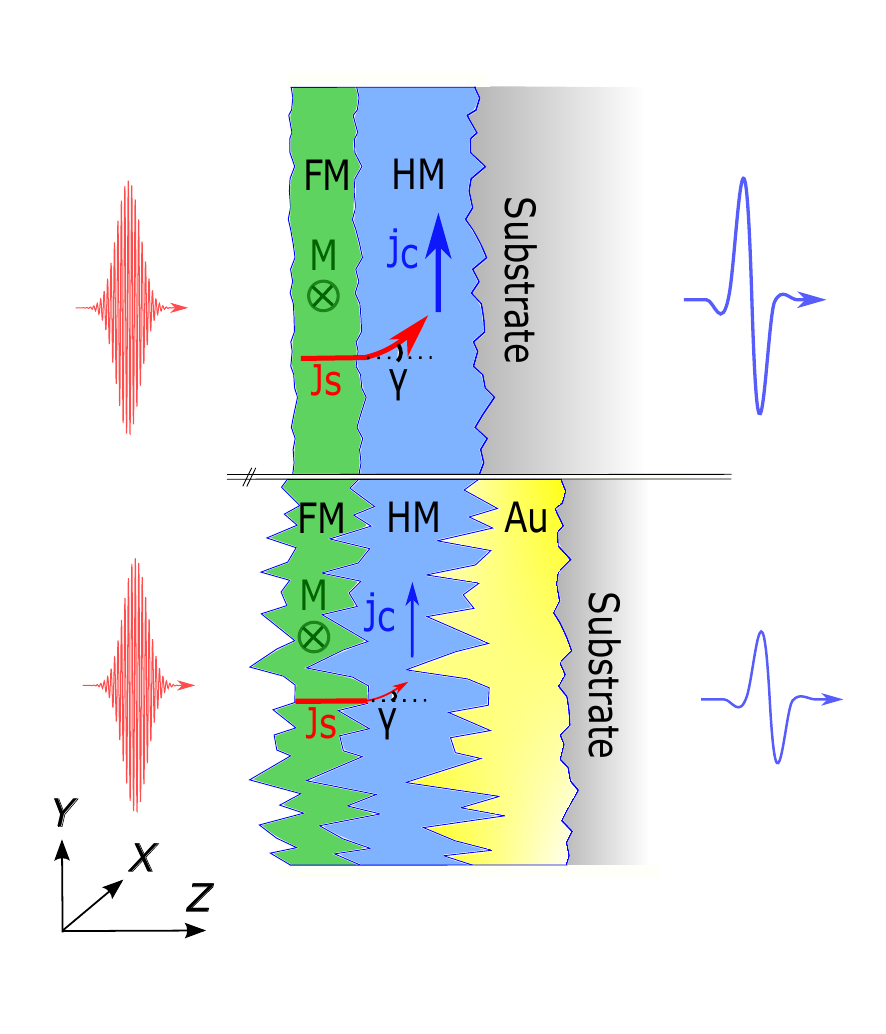}
\caption{THz emission from FM$|$HM samples with variable buffer layer roughness. A fs laser pulse excites a ferromagnetic layer and injects a spin-current pulse $j_s$ into a heavy metal layer. $j_s$ is affected by the FM$|$HM interface spin transparency $t_s$. Two extremal cases are shown: without an Au buffer layer (smooth interface) and with a thick Au buffer layer creating a very rough interface.}
\label{fig1}
\end{figure}

Our general approach to generate ultrafast spin-current flowing through an interface of variable roughness is shown in Fig.~\ref{fig1}. The sample stack consists of FM$|$HM$|$Au($d$), where $d$ indicates the variable thickness of the Au buffer layer, which affects its roughness and is translated also in the FM$|$HM interface roughness \cite{Verbeno2023}. First, the femtosecond laser pulse excites the sample stack and transfers a fraction of the incident optical power $A_{\text{FM}}(d)$ into the FM layer, depending on the thickness $d$, due to propagation of light in layered media. This generates an ultrafast spin voltage in FM \cite{Seifert2022, Rouzegar2022} and consequently an ultrafast spin current pulse $j_s(d)$ traversing through the interface with spin transparency $t_s$ into HM layer. The amplitude of spin voltage and $j_s$ is assumed to scale with $A_{\text{FM}}$ and $t_s$ \cite{Jechumtal2024}. Second, the out-of-plane spin-current pulse in HM is converted into the in-plane charge-current $j_c \propto \gamma_{\text{SH}} j_s$ via the ISHE, where $ \gamma_{\text{SH}} $ is the spin-Hall angle in HM. Finally, the $j_c$ emits a THz pulse with electric field $E_{\text{THz}}$ right behind the sample \cite{Seifert2016, Seifert2018}:
\begin{equation}
\begin{split}
E_{\text{THz}}(d) = \frac{e Z_0}{n_1 + n_2 + Z_0 G(d)} j_c(d) \propto  \\
t_s(d) A_{\text{FM}}(d)Z(d) M_{\text{FM}}\gamma_{\text{SH}}\lambda_{\text{HM}}\tanh{\frac{d_{\text{HM}}}{2\lambda_{\text{HM}}}}
\end{split}
\label{eq1}
\end{equation}
where the first equality means that the emitted THz field relates to the in-plane charge current $j_c$ via the generalized impedance of the sample, where $e$ is the elemental charge, $Z_0 = 1/(\epsilon_0 c) \approx 377\,\Omega$ is the vacuum impedance, $n_1, n_2$ are refractive indices of sample silicon substrate ($\approx 3.4$) and the surrounding atmosphere ($\approx 1$), and $G(d)$ is the total conductance of the whole sample stack. The second relation corresponds to the mechanism of $j_c$ generation and includes magnetization of FM layer $M_{\text{FM}}$ and the factor $\lambda_{\text{HM}} \tanh \frac{d_{\text{HM}}}{2\lambda_{\text{HM}}}$ coming from the integration of $j_s$ in heavy metal with thickness $d_{\text{HM}}$ and spin current relaxation length $\lambda_{\text{HM}}$ \cite{Seifert2018}. Relation $\frac{e Z_0}{n_1 + n_2 + Z_0 G(d)} = Z(d)$ for generalized impedance was used. From Eq.~\ref{eq1}, we see that the spin transparency $t_s(d)$ is directly proportional to $E_{\text{THz}}(d)$ and we can observe solely the role of buffer Au layer thickness $d$ on $t_s(d)$ just by renormalizing $E_{\text{THz}}(d)$ on changing impedance $Z(d)$ and optical energy absorbed in FM $A_{\text{FM}}(d)$.

In the experiment, we measure the electro-optical THz signal $EOS_{\text{THz}}$, which is related to $E_{\text{THz}}$ at the sample surface via the convolution with the setup function: $EOS_{\text{THz}}(t)=[H*E_{\text{THz}}](t)$. Owing to the fact that $H(t)$ is not changing within our experiment, for spectrally identical signals, we can relate the strength of $EOS_{\text{THz}}$ with $t_s(d)$ in the following way:
\begin{equation}
    S_{\text{THz}}(d)\propto E_{\text{THz}}(d) \propto t_s(d)A_{\text{FM}}(d) Z(d)
    \label{eq2}
\end{equation}

\subsection{Sample set and its growth}

The impact of Au layer thickness $d$ and the induced roughness of the FM$|$HM interface is characterized by an atomic force microscope (AFM) on a calibration sample without the top FM layer. Two cases shown in Fig.~\ref{fig1} schematically represent samples without an Au layer and with a relatively thick and rough Au buffer layer.

The influence of interfacial roughness on spin transparency was studied using two sets of samples. THz measurements were performed on Au(5)$|$Co(2)$|$Pt(5)$|$Au($d$)$|$Si, with the thickness of the gold buffer layer ranging from $d = $ 0 to $d = $ 20~nm. Platinum is chosen as the HM because of the very efficient spin-to-charge conversion by the strong (I)SHE with $\gamma_{\text{SH}} \sim 10$\% \cite{Sinova2015, Tao2018}, which causes the conversion to be dominated solely in Platinum. The platinum crystalline quality and consequently its spin hall angle $\gamma_{\text{SH}}$ is assumed to be the same across the whole sample set, due to constant deposition conditions. Part of the sample area was shielded in the sputtering process and reveals the bare substrate, which allows us to easily measure a reference THz transmission when evaluating  $Z(d)$ \cite{Nadvornik2021}. The top Au layer, serving as a capping layer to prevent cobalt oxidation, was nominally identical across all prepared multilayers and therefore did not contribute to any differences across the sample set. The second set of samples served as a characterization tool for the Co$|$Pt$|$Si interface in the first set and consisted of Pt(5)$|$Au($d$) stacks with the same buffer layer thicknesses, namely $d = 0, 1, 3, 4, 5, 6, 8, 10, 15$, and $20$~nm. 

All samples were prepared by magnetron sputtering in a Bestec GmbH ultrahigh-vacuum system with a base pressure below $10^{-8}$~mbar. The sputtering process was performed at room temperature, with working pressures of $3.6 \cdot 10^{-3}$~mbar for Co and $5 \cdot 10^{-3}$~mbar for all other materials, using argon gas. The substrates were rotated at 30 RPM during deposition to achieve a homogeneous coating. Naturally oxidized, undoped silicon (100) substrates were used, cleaned ex situ in an acetone and isopropyl alcohol bath prior to sputtering. As the set of samples intended for interface characterization did not have any capping applied, the measurements were performed immediately after sample preparation. Although the top platinum layer might experience some oxidation, all samples were characterized within the same time frame after preparation; therefore, a similar process across all samples is expected, potentially resulting in a systematic offset in the obtained data.

\subsection{THz experimental setup}

In the THz emission experiments, the sample is excited by a train of ultrashort laser pulses (wavelength 1030~nm, duration $\sim$170~fs, repetition rate 10~kHz, energy per pulse 10~$\mu$J) from a Pharos laser system. The collimated pump beam has a spot with a full width at half maximum (FWHM) of the intensity of $\sim$3~mm on the sample. The magnetization $M_{\text{FM}}$ of the Co layer is controlled by an external magnetic field of $\sim$300~mT. The emitted THz electric field $E_{\text{THz}}$ propagates through a set of parabolic mirrors in ambient atmosphere and is detected as an electro-optical (EO) signal $S_{\text{THz}}$ via EO sampling in a 2~mm thick GaP(110) crystal by using linearly polarized probe pulses split from the pump beam. The spectral bandwith of the experimental setup is typically from 0.1 to 2.5~THz, as shown in Fig. \ref{fig2}(b).

\subsection{Interface roughness characterization}

Interfacial roughness and surface topography of the Pt(5)$|$Au($d$) set of samples were characterized using atomic force microscopy in contact mode. Three $1\times1$~$\mu$m$^2$ scans at random spots were evaluated for all samples with respect to surface root-mean-squared (RMS) roughness, the total integrated surface area, and grain size using the program Gwyddion \cite{Gwyddion12}. After subtracting the measurement background, the RMS was evaluated using the statistical quantities function. Grain features on the measured topography were segmented using Otsu's method, and the average equivalent disk radius (EDR) was extracted as a measure of grain size. The presented data are the averages of the forward and backward scans across all measured spots on a particular sample.

The topography measurements were performed on a Park Systems NX 10 AFM system using a silicon AFM probe (PPP-CONTSCR; Nanosensors) with a spring constant $k=0.02$~N·m$^{-1}$ (measured by the thermal noise method) and a nominal tip radius $R=10$~nm. The scan rate was set to $0.2$~Hz. The scan size was 512 pixels per line, providing a lateral resolution of $\sim 2$~nm.

\section{Results}

\begin{figure*}[t]
\centering
\includegraphics[width=0.95\textwidth]{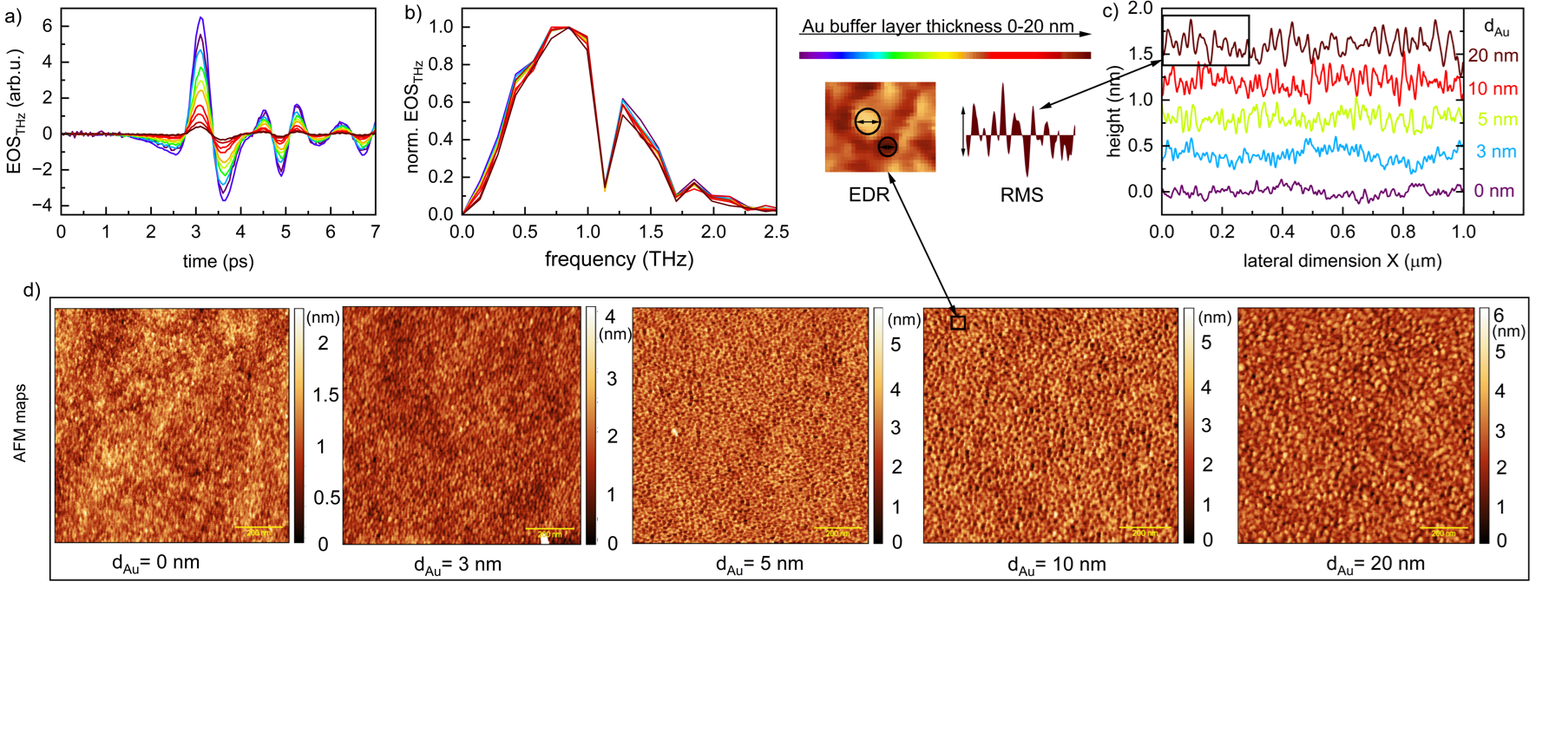}
\caption{(a) Raw THz signals measured by electro-optical sampling from a set of samples Au(5)$|$Co(2)$|$Pt(5)$|$Au($d$). (b) Spectra of THz waveforms from (a). Spectral dips come from water vapor absorption. The thickness of the Au buffer layer for the given sample is represented by the line colour shown on the colour bar. (c) Representative topography profiles of the Pt(5)$|$Au($d$) set of samples; note that the vertical axis is on the nanometer scale, while the horizontal axis is on the micrometer scale. (d) Selected Atomic Force Microscopy scans of the Pt(5)$|$Au($d$) set of samples. The topography maps were measured in contact mode, revealing the building of a columnar-like structure with increasing thickness of the Au buffer layer. The figure includes a schematic sketch which shows that the RMS of interface roughness represents the vertical roughness, whereas the EDR represents rather the lateral roughness.}
\label{fig2}
\end{figure*}

Typical THz-emission waveforms $S_{\text{THz}}(d)$ from the Co$|$Pt$|$Au($d$) samples are shown in Fig.~\ref{fig2}(a). As $d$ increases from 0 to 20~nm, the overall signal amplitude increases for the very first sample with a non-zero Au layer ($d = 1$~nm) and then decreases by roughly a factor of 10 for the thickest sample ($d = 20$~nm). Within the experimental precision, the spectral characteristics of THz waveforms remain unchanged as shown in \ref{fig2}(b). Corresponding topography profiles of the Pt interface for 5 representative samples are shown in Fig. \ref{fig2}(c), and the AFM topography maps are shown in Fig. \ref{fig2}(d).

\begin{figure}[t]
\centering
\includegraphics[width=\columnwidth]{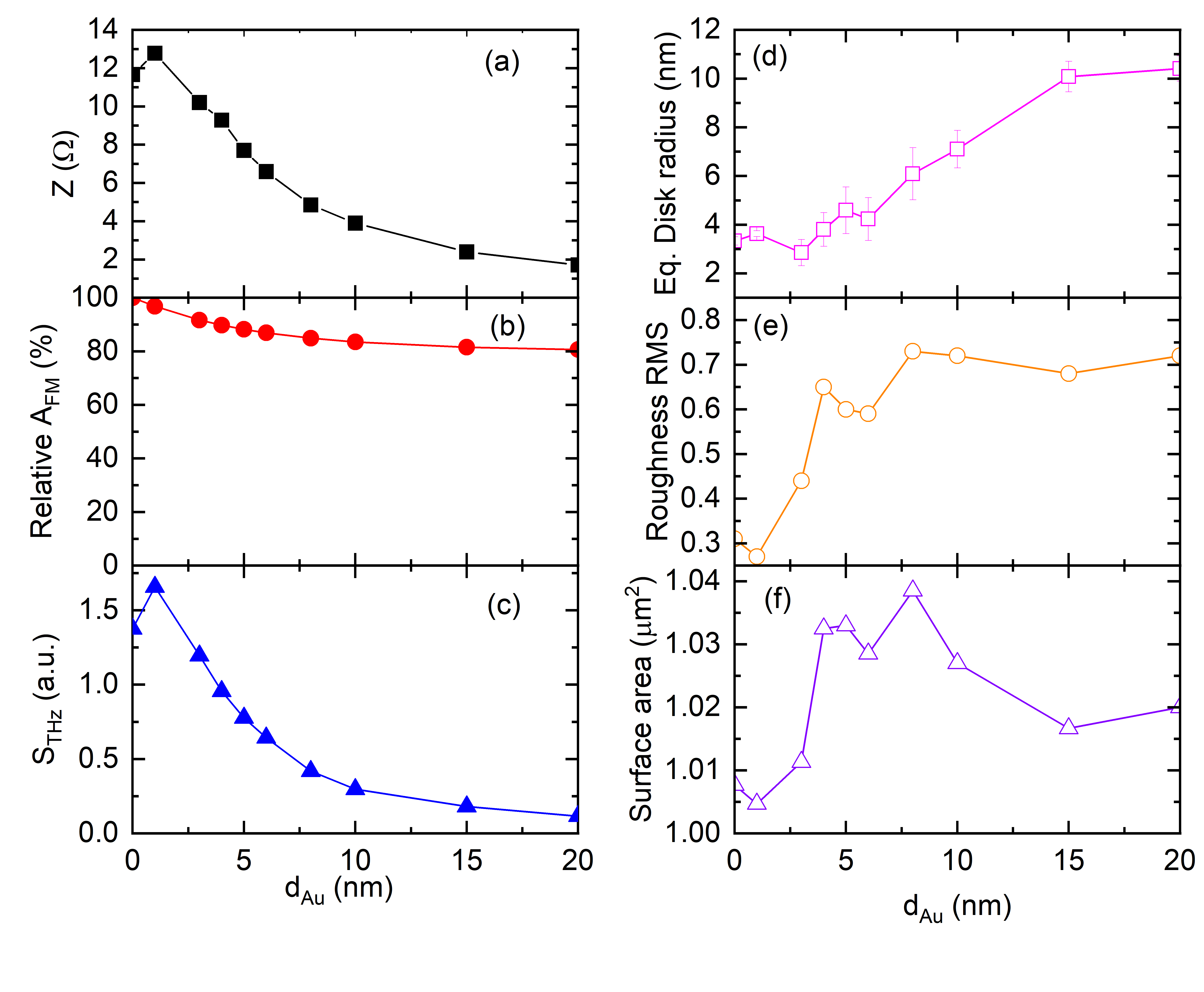}
\caption{Sample properties as a function of the Au buffer layer thickness $d$. (a) Measured total sample impedance $Z(d)$. (b) Calculated relative change of energy absorbed in the FM layer $A_{\text{FM}}(d)$ normalized to the value at $d=0$. (c) Magnitude of THz signal $S_{\text{THz}}$. (d) Interface roughness EDR (lateral roughness). (e) RMS (vertical roughness) of the AFM map. (f) Interface surface area as a function of Au thickness $d$.}
\label{fig3}
\end{figure}

In more detail, the total THz impedance $Z(d)$ of samples is shown in Fig.~\ref{fig3}(a). The normalized optical power absorbed in the Co layer as a function of $d$, $A_{\text{FM}}(d)$, is shown in Fig.~\ref{fig3}(b). The absorbed power was calculated using ANSYS Numerical FDTD solver, taking into account the whole multilayer stack. The absorption changes are mainly driven by the back-reflected wave from the substrate. As the thickness of the Au buffer layer increases, this layer starts to absorb the light more, decreasing the power of back-reflected light, which can be absorbed in Co. The total strength of $EOS_{\text{THz}}(d)$, calculated as the root mean square of THz waveforms, $S_{\text{THz}}(d)$, is shown in Fig.~\ref{fig3}(c), clearly reflecting the non-monotonic trend observed in the raw amplitudes.

The changing morphology of the Co$|$Pt interface was evaluated from AFM images and profiles shown in Fig.~\ref{fig2} (c) and (d). In the absence of a buffer layer, the interface is solely determined by the roughness of the substrate and its natural oxidation state, which is relatively non-uniform \cite{Morita90}. At low buffer thicknesses, the growth of the Au layer follows the Volmer--Weber mode, where small island-like structures grow on the substrate due to Au atoms binding more strongly to each other than to the substrate, producing a strained discontinuous layer \cite{Abadias2015, Liang2022, Malinsky2012}. With intermediate Au thicknesses, these small islands merge and form a continuous layer, albeit still strained, with distinct island- or columnar-like growth that becomes more pronounced at higher thicknesses. Upon capping this layer with platinum, the resulting top surface is continuous, but the overall roughness is inherited from the underlying Au layer.

We evaluated the mean EDR [Fig.~\ref{fig3}(d)] describing the "lateral" roughness by the size of grains, the RMS of surface roughness [Fig.~\ref{fig3}(e)] characterizing the "vertical" roughness of the interface, and the total integrated surface area [Fig.~\ref{fig3}(f)]. For an intuitive understanding of roughness, RMS, and EDR, see the schematic sketch included in Fig.~\ref{fig2}. We observe a clear positive correlation of both EDR and RMS with the buffer layer thickness $d$. In contrast, the total surface area of the interface exhibits a more complex, non-monotonic dependence on $d$. Using the relation derived from Eq.~\ref{eq2}, we extracted the spin transparency $t_s$ from $S_{\text{THz}}$ by accounting for the changes in impedance $Z(d)$ and absorption within the Co layer $A_{\text{FM}}(d)$, finding a $\sim$30\% decrease across the studied range.

\begin{figure}[t]
\centering
\includegraphics[width=\columnwidth]{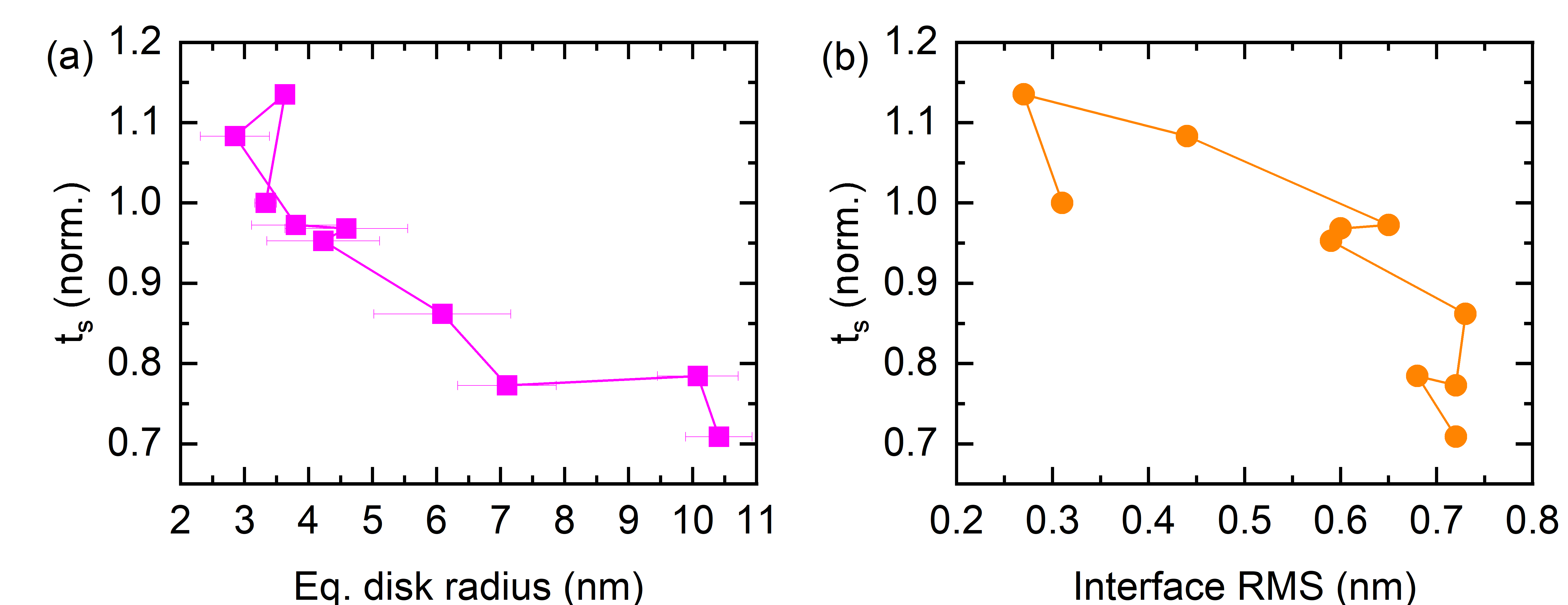}
\caption{Spin transparency $t_s$ extracted from $S_{\text{THz}}$ as a function of surface roughness parameters (a) EDR (lateral roughness) and (b) RMS (vertical roughness). Both cases show a clear negative correlation, with $t_s$ decreasing by $\sim$30\% across the studied range.}
\label{fig4}
\end{figure}

The correlation of $t_s$ with the surface roughness parameters is shown in Fig.~\ref{fig4}. For both parameters, the EDR [Fig.~\ref{fig4}(a)] and RMS [Fig.~\ref{fig4}(b)], we observe a clear negative correlation. Interestingly, $t_s$ for $d = 1$~nm slightly exceeds that of the sample without any Au buffer layer. This initial enhancement, which is also responsible for the peak in the raw THz signal, can be attributed to the very early stage of Au nucleation. At this stage, sub-nanometer Au islands have not yet percolated into a continuous layer (and thus do not decrease the total stack impedance), but they effectively promote improved wetting and adhesion of the subsequently deposited Pt layer.

\section{Discussion}

The observed decrease in spin transparency $t_s$ with increasing interface roughness can be understood by considering the increased spin-orbit interaction (SOI) and scattering at the interface \cite{Amin2016}. One possible mechanism for such a decrease could be enhanced intermixing at the Co$|$Pt interface induced by the rougher underlying morphology, which might create a magnetically dead layer or other conversion channels for THz radiation \cite{Gueckstock2021}. However, this scenario is excluded based on previously published results on nominally identical samples \cite{Verbeno2023}, where the saturation magnetization $M_s$ of the Co layer remained constant across the full range of Au buffer layer thicknesses. Given that the Co layer is only 2 nm thick, any significant interfacial intermixing or dead layer would reduce the total magnetization. The invariance of $M_s$ with $d$ provides direct evidence that interface intermixing is negligible.

When plotted against the RMS roughness, $t_s$ shows a gradual decrease that approximately follows the increase of the RMS parameter. However, the correlation is not strictly linear, which can be attributed to the non-monotonic changes of RMS at low Au thicknesses. A more systematic correlation is observed between $t_s(d)$ and the EDR, corresponding to the mean grain size. The grain size exhibits a sigmoidal-like increase with $d$, featuring a rapid change around $d \sim 6$--$8$~nm that coincides with the transition from non-uniform coalescence to established columnar growth. This suggests that the lateral size of the grains matters more for spin transparency than the vertical roughness alone. Larger grain sizes correspond to longer-range lateral roughness of the interface, which may enhance spin-flip scattering at the Co$|$Pt boundary due to locally varying interface normal directions and the associated variation of the spin-mixing conductance \cite{Zhu2019, Pham2018}.

The relatively modest decrease of $t_s$ across the studied range can be rationalized by comparing the relevant length scales. The RMS roughness reaches values between $0.3$ and $0.8$~nm, which is smaller than the spin diffusion length in Pt ($\lambda_{\text{Pt}} \approx 1$--$3$~nm \cite{Rojas2014, Tao2018, Bass2007}). In this regime, the spin current traversing the interface is only partially affected by the lateral roughness (characterized via EDR), as the spin-coherence volume effectively averages over the local interface normal variations within each grain. A more dramatic suppression of $t_s$ could be expected only when the RMS amplitude significantly exceeds $\lambda_{\text{HM}}$.

The spectral dependence of the THz waveform doesn't depend on the interface roughness [Fig. ~\ref{fig2} (b)],  which further supports the interpretation that the interface roughness modifies the overall magnitude of spin current transmission rather than introducing frequency-dependent scattering channels. This is consistent with the fact that the relevant roughness length scales (1--10~nm) are orders of magnitude smaller than the THz wavelength. Furthermore, the spin-current pulse traverses the interface on a timescale shorter than the characteristic time associated with lateral diffusion across individual grains.

It is instructive to compare our results with quasi-static spin-pumping experiments. In the structure of ferromagnetic insulator (FMI) and heavy metal Fe$_3$O$_4$$|$Pt, interface roughness was reported to increase the spin-mixing conductance by approximately 40\% \cite{Pham2018}. The opposite trend observed in our metallic Co$|$Pt system reflects the distinct spin transport mechanisms: in insulating Fe$_3$O$_4$$|$Pt, roughness enlarges the effective magnon-electron coupling area, while in the metallic Co$|$Pt system, the spin current is carried by hot electrons whose spin-flip scattering probability increases with interface roughness. 

Our results can be directly compared to a recent ST-FMR study  \cite{Gu26}, in which the Pt$|$NiFe interfacial roughness was tuned by in-situ Ar plasma treatment of the Pt surface before NiFe deposition. There, the spin transparency decreased significantly faster with increasing roughness than in our THz experiment: only a $\sim$25\% change in interfacial roughness reduced the initial interface spin transparency by roughly the same factor as in our case, with a factor-of-three change in interface RMS. We attribute this difference to the disparity of relevant length scales between the two regimes. At GHz, the spin-pumping signal integrates over many precession cycles, sampling the lateral grain morphology and accumulating scattering contributions across the full interface. In contrast, the sub-picosecond spin-current pulse generated by THz emission traverses the interface before lateral diffusion across an individual grain can occur, effectively averaging only over a sub-grain region whose local roughness is far smaller than the global RMS. In other words, on picosecond timescales, the spin transparency is less sensitive to interface roughness than in the quasi-static regime.

Despite this difference in slope, the total magnitude of the roughness-induced change in our experiment is of a similar order to that reported in GHz and DC spin-pumping studies \cite{Gu26, Pham2018, Zhu2019}, reflecting the fact that our wider roughness window compensates for the shallower dependence. This suggests that the ultrafast character of the spin-current pulse does not introduce additional roughness-sensitive scattering channels beyond those already present at lower frequencies.

\section{Conclusion}

In summary, we employed terahertz emission spectroscopy to directly extract the change of ultrafast spin transparency $t_s$ across a set of samples with varied interface roughness, controlled by the gold buffer layer thickness. We observed a clear negative correlation of $t_s$ with both the interface roughness RMS and the equivalent disk radius (EDR) of the surface grains. We conclude that the decreased $t_s$ is primarily driven by enhanced spin-orbit interaction and spin-flip scattering at the interface, rather than material intermixing. The lateral grain morphology appears to play a crucial role, with the transition to a columnar growth regime at $d \sim 6-8$~nm marking a significant change in the interface properties. Crucially, our results demonstrate that even when the interface morphology is significantly altered, the spin transparency remains remarkably high. This robustness is highly promising for future applications, indicating that highly efficient spintronic THz emitters can be reliably achieved in large-area, mass-produced fabrication processes where perfect interfacial smoothness is difficult to maintain.

\section{Acknowledgements}
The authors acknowledge funding by the Ministry of Education, Youth and Sports of the Czech Republic through the OP JAK call Excellent Research
(TERAFIT Project No. CZ.02.01.01/00/22-008/0004594 and FerrMion Project No. CZ.02.01.01/00/22-008/0004591) and CzechNanoLab Research Infrastructure (LM2023051), the Czech Science Foundation through projects TA CR (Grant No. 24–11361S), the Grant Agency of the Charles University (Grant No. 120324 and SVV–2025–260836) and the European Union’s Horizon Europe programme through project Tera-MaRs (No. 101211111).

\section{Data Availability}
The data that support the findings of this article are openly available \cite{DataZenodo}.

\setcounter{figure}{0}
\renewcommand{\thefigure}{S\arabic{figure}}

\bibliographystyle{apsrev4-2}
\bibliography{refs}

\end{document}